\documentclass[fleqn,usenatbib]{mnras}

\usepackage[T1]{fontenc}
\usepackage{ulem}
\usepackage{soul}

\DeclareRobustCommand{\VAN}[3]{#2}
\let\VANthebibliography\thebibliography
\def\thebibliography{\DeclareRobustCommand{\VAN}[3]{##3}\VANthebibliography}

\usepackage{graphicx}	% Including figure files
\usepackage{amsmath}	% Advanced maths commands
\usepackage{amssymb}	% Extra maths symbols
\usepackage{caption}
\usepackage{subcaption}
\usepackage{mwe}
\usepackage[dvipsnames]{xcolor}
\usepackage{hyperref}
\newcommand{\referee}{}

\usepackage{ulem}

\title[Magnetic braking and the evolution of LMXBs]{The impact of different magnetic braking prescriptions on the evolution of LMXBs}

 \author[M. Echeveste, M. L. Novarino, O. G. Benvenuto \& M. A. De Vito]
 {
 M. Echeveste\thanks{Fellow of the Consejo Nacional de Investigaciones Cient\'{\i}ficas y T\'ecnicas (CONICET). Email: mecheveste@fcaglp.unlp.edu.ar},
 M. L. Novarino\thanks{Fellow of the CONICET. Email: leonova@fcaglp.unlp.edu.ar},
 O. G. Benvenuto\thanks{Member of  the Carrera del Investigador
 Cient\'{\i}fico, Comisi\'on de  Investigaciones Cient\'{\i}ficas de la  Provincia de Buenos Aires (CIC). Email: obenvenu@fcaglp.unlp.edu.ar},
 M. A. De  Vito\thanks{Member of  the Carrera del Investigador
 Cient\'{\i}fico of CONICET. Email: adevito@fcaglp.unlp.edu.ar}
 \\
  Instituto de Astrof\'{\i}sica de La Plata, IALP, CCT-CONICET-UNLP, La Plata, Argentina and\\
 Facultad de Ciencias Astron\'omicas y Geof\'{\i}sicas, Universidad
 Nacional de La Plata (UNLP),\\ Paseo del Bosque S/N, B1900FWA, La Plata,
 Argentina}

\date{Accepted XXX. Received YYY; in original form ZZZ}

\pubyear{2023}
\usepackage{newtxtext,newtxmath}
\begin{document}
\label{firstpage}
\pagerange{\pageref{firstpage}--\pageref{lastpage}}
\maketitle

\begin{abstract}
We revisit the evolution of low-mass close binary systems \referee {under} different magnetic braking (MB) prescriptions. \referee{We study binaries with a neutron star accretor. During mass transfer episodes, these systems emit X-rays and are known as Low Mass X-ray Binaries (LMXBs). When mass transfer stops, they can be observed as binary pulsars. Additionally, some of these systems can experience mass transfer while having orbital periods of less than 1~hr, thus evolving into ultracompact X-ray binaries (UCXBs). }The evolution of LMXBs depends on their capability to lose angular momentum and maintain stable mass transfer. Among the angular momentum loss mechanisms, MB is one important, and still uncertain phenomenon. The standard MB prescription faces some problems when calculating LMXB evolution, leading to, e.g., a fine-tuning problem in the formation of UCXBs. Recent studies proposed new MB prescriptions, yielding diverse outcomes. Here, we investigate the effects of three novel MB prescriptions on the evolution of LMXBs using our stellar code. We found that all MB prescriptions considered allow the formation of binaries with orbital periods spanning from less than one hour to more than tens of days. Remarkably, our results enable the occurrence of wide systems even for the MB law that causes the strongest angular momentum losses and very high mass transfer rates. \referee{We found that models computed with the strongest MB prescription reach the UCXB state starting from a wider initial orbital period interval.} Finally, we discuss and compare our results with observations and previous studies performed on this topic. 
\end{abstract}

\begin{keywords}
binaries (including multiple): close -- stars: evolution --  pulsars: general
\end{keywords}

%%%%%%%%%%%%%%%%%%%%%%%%%%%%%%%%%%%%%%%%%%%%%%%%%%

%%%%%%%%%%%%%%%%% BODY OF PAPER %%%%%%%%%%%%%%%%%%

\section{Introduction}
\label{intro}

Low-mass X-ray binaries (LMXBs) are systems composed of a low-mass donor star that transfers mass to a compact star, either a neutron star (NS) or a black hole, via Roche-lobe overflow (RLOF) (see \citealt{1993ARA&A..31...93V}; \citealt{2010csxs.book..623T} for a review). A few hundred LMXBs were detected in our galaxy (\citealt{2007A&A...469..807L}) and in external galaxies (\citealt{2003ApJ...585..756J}; \citealt{2008ApJ...689..983H}; \citealt{2016ApJ...829...20W}). Here we study the case of LMXBs with NS acretors. 

If the orbit is not too wide, the donor star fills its Roche Lobe and transfers mass and angular momentum to the NS, making it spin up in a process known as ``recycling''. When mass transfer ceases, the system can be observed as a binary pulsar (BP) (\citealt{1991PhR...203....1B}; \citealt{2010csxs.book..623T}). The companion star is then typically a Helium White Dwarf (He~WD), a Carbon-Oxygen White Dwarf (CO~WD), or a low-mass dwarf, depending on the initial orbital period and mass of the donor (\citealt{2011ASPC..447..285T}; \citealt{2013IAUS..291..127R}; \citealt{2013ApJ...775...27C}). Evolutionary paths from LMXBs to BPs with He~WD companions have been broadly studied (\citealt{1999A&A...350..928T}; \citealt{2002ApJ...565.1107P}; \citealt{2014A&A...571A..45I}). \referee{Additionally, binary millisecond pulsars (BMPs) with Extremely Low-Mass (ELM) WDs ($M_{\rm WD} \lesssim 0.20~M_{\odot}$) and short orbital periods ($P_{\rm orb} \simeq  2 - 9$ hr) can start mass transfer once again, evolving into Ultra-Compact X-ray Binaries (UCXBs). Studying the efficiency of the MB laws on the production of UCXBs and BMPs with ELM WDs is interesting, as these objects are candidates to be gravitational wave sources. They could be detected in the mHz frequency band by different missions, e.g. LISA (\citealt{2017arXiv170200786A}), TianQin (\citealt{Luo_2016}), and Taiji (\citealt{2018arXiv180709495R}).} 

There are many angular momentum loss mechanisms acting on the evolution of these binaries. Usually, the most considered are magnetic braking (MB), gravitational radiation, and mass loss. Among all these, MB is particularly uncertain. The most used MB law among the simulations of low-mass binaries is that derived by \citet{1981A&A...100L...7V}; \citet{1983ApJ...275..713R}, which is based on the empirical Skumanich law \citep{1972ApJ...171..565S}. In this law, the angular momentum loss $\dot{J}_{\rm MB}$ scales with the donor mass $M_2$, the donor radius $R_2$, and angular velocity $\omega_2$. It was built to describe main-sequence stars comparable to our Sun. Nevertheless, as \citet{2019MNRAS.483.5595V} pointed out, another kind of donor may have an increased mass loss rate with the stellar wind and a magnetic field strength that does not scale directly with the angular velocity of the star. In any case, these may play a role in the MB prescriptions (\citealt{1968MNRAS.138..359M}; \citealt{1987MNRAS.226...57M}; \citealt{1988ApJ...333..236K}). 

The prescription based on the Skumanich law faces some problems when calculating LMXBs evolution. Notably, the orbital period distribution of BMPs descending from LMXBs disagrees with observations, especially for periods between 0.1 and 10~d (\citealt{2003ApJ...597.1036P}; \citealt{2014A&A...571A..45I}; \citealt{2015ApJ...809...99S})\referee{: the number of objects observed in this orbital period range is far higher than theoretical expectations}. Besides, the mass accretion rate inferred by observations is about one order of magnitude higher than those calculated using this law (\citealt{2002ApJ...565.1107P}; \citealt{2003ApJ...597.1036P}; \citealt{2015ApJ...809...99S}; \citealt{2016MNRAS.456..263P}; \citealt{2019MNRAS.483.5595V}). Furthermore, it was also found that UCXBs are very difficult to reproduce under this MB law: only systems within a very narrow range of initial orbital period and donor mass can evolve into UCXBs (fine-tuning problem) (\citealt{2005A&A...431..647V}; \citealt{2005A&A...440..973V}; \citealt{2014A&A...571A..45I};  \citealt{2019BAAA...61...87E}). \referee{In addition, there is a fine-tuning problem in the formation of BMSPs with ELM WDs descending form LMXBs (\citealt{2005A&A...431..647V}; \citealt{2005A&A...440..973V}; \citealt{2014A&A...571A..45I}).}

%only LMXBs in a very narrow range of initial orbital periods can reproduce BMSPs with ELM WDs (\citealt{2005A&A...431..647V}; \citealt{2005A&A...440..973V}; \citealt{2014A&A...571A..45I})}

Motivated by the points mentioned above, \citet{2019MNRAS.483.5595V} investigated modifications to the Skumanich law and studied their implications in the mass accretion rate problem. They presented three new prescriptions for MB, named Convection-boosted (MB2), Intermediate (MB3), and Wind-boosted (MB4). These prescriptions include scaling of the magnetic field strength with the convective turnover time and the wind mass-loss rate. In that work, they also argue that the MB4 prescription may need additional effects to dampen the angular momentum loss since it results in mass transfer rates that exceed $1~M_\odot \rm{yr^{-1}}$. Later, \citet{2019ApJ...886L..31V} presented a fifth MB prescription, named Convection And Rotation-Boosted (CARB~MB). This is obtained through a self-consistent deduction that includes the effect of the stellar rotation on the Alfvèn radius and the magnetic field dependence on the convective turnover time.

\referee{According to \citet{2019ApJ...886L..31V}, all LMXBs of interest can be reproduced under the MB3 prescription when matching three observed parameters (e.g., orbital period, mass transfer rate [hereafter MTR], and mass ratio). Nevertheless, if the effective temperature is also considered, they could no longer reproduce the system Sco X-1 with this MB law. Alternatively, when CARB MB is used, they notice that all persistent LMXBs can be reproduced. Therefore, they recommend its use instead of the Skumanich MB to model galactic and extragalactic LMXBs with NSs.}

\referee{This paper aims to explore the effects of the Skumanich, Convection-boosted, Intermediate, and  Convection and Rotation-boosted MB prescriptions on the evolution of LMXBs in a global way. Additionally, we focus on the capability of these prescriptions in producing ELM WDs BMPs, and UCXBs}. For such purpose, we employ our binary stellar evolution code, which is fully independent of the public domain {\tt MESA} code \citep{2015ApJS..220...15P} that has been used by other researchers in similar investigations.  % \ogb{Para que citarlo tantas veces?}. \sout{\citep{2013ApJS..208....4P, 2015ApJS..220...15P, 2018ApJS..234...34P, 2019ApJS..243...10P}}.

The remainder of this paper is organised as follows. In Section \ref{sec:code} we describe the numerical methods and physical assumptions made up in our code. In Section \ref{sec:results} we present the results obtained from our simulations and compare them with observations. In Section~\ref{sec:comparison} we compare our results with those obtained in previous studies. Finally,  in Section~\ref{sec:Disc-Concl} we summarise our main findings and make some concluding remarks.

%-----------------------------------------------------------------------------
\section{Evolutionary Code and Binary Models} \label{sec:code}

We calculate \referee{a set of numerical models with an updated version of the binary evolutionary code presented in \citet{2003MNRAS.342...50B}. When the components of the binary system are detached, our code employs the standard Henyey technique. When the configuration is semi-detached, it incorporates the determination of the MTR from Roche lobe overflow as an additional variable to be iterated consistently, as will be described in Subsection~\ref{subsec:MdotRLOF}} The \referee{treatment of the} mass transfer is included as in, e.g., \citet{2002ApJ...565.1107P}, by introducing two free parameters\referee{:} $\alpha$ and $\beta$. $\beta$ is the fraction of mass lost by the donor that may be accreted by the compact object ($\dot{M}_{1}= -\beta \dot{M}_{2}$\referee{; hereafter, we adopt label $1$ to refer to the compact object and $2$ to the donor star}). As standard, we set an upper limit to $\dot{M}_{1}$ imposed by the Eddington rate ($\dot{M}_{\rm Edd}= 2 \times 10^{-8}\; M_{\odot}/\rm{y}$), \referee{then} $\dot{M}_{1}= \min\big( -\beta \dot{M}_{2}, \dot{M}_{\rm Edd}\big)$. If the compact star \referee{cannot} retain all the mass transferred by the donor, it is lost from the system carrying away angular momentum with a rate $\dot{J}_{\rm ML}$. The parameter $\alpha$ is introduced to describe this loss in units of the specific angular momentum of the compact component $j_{1}$ as $\dot{J}_{\rm ML}= \alpha (\dot{M}_{1}+\dot{M}_{2})\; j_{1}$. Gravitational wave radiation is described as in \citet{1975ctf..book.....L}. MB is modeled under different prescriptions (see Subsection~\ref{subsec:Mbraking}). We consider spin-orbit coupling in a way similar to that performed in \citet{2021MNRAS.508.3812N} (see Subsection~\ref{subsec:tides}). Besides, we include the wind mass loss of the donor following \citet{1975psae.book..229R}. \referee{Also, we consider the effect of the binding energy of the NS on the total amount of accreted matter, calculated as in \citet{2011MNRAS.413L..47B}}. 

\subsection{Mass transfer from RLOF} \label{subsec:MdotRLOF}

Our code is based on a modification of the Henyey technique presented in \citet{1967MComP...7..129K} to solve the set of difference equations of stellar evolution together with MTR from RLOF. MTR is included as an additional variable in the Henyey method and calculated simultaneously with the solution of the equations of stellar structure. \referee{Therefore, MTR is calculated in a self-consistent way}: during mass transfer episodes, the code finds the MTR in a fully implicit way employing a {\it single} iterative procedure. \referee{This type of algorithm exhibits robust numerical stability, as was previously discussed in \citet{2006A&A...445..647B}}. This procedure is different from that employed by {\tt MESA} and other researchers (e.g. \citealt{2002ApJ...565.1107P}; \citealt{2008ApJ...688.1235M}; \citealt{2011ApJ...732...70L}). They apply a {\it double} iterative procedure in constructing a stellar model: an MTR \referee{value} is estimated and the stellar structure is relaxed for this MTR. Then, with this new structure, a new MTR is computed and the stellar structure is relaxed again. This is repeated until consistency is achieved. 

We calculate MTR ($\dot{M}_{\rm RLOF}$) using either the {\it Ritter} \citep{1988A&A...202...93R} or the {\it Kolb-Ritter} \citep{1990A&A...236..385K} schemes. \referee{Applying one or the other scheme depends upon the relation between the radius of the donor star, $R_2$, and the equivalent radius}\footnote{\referee{This is the radius of a sphere that has the same volume as the Roche Lobe.}} of its Roche Lobe, $R_{\rm L}$.

The {\it Ritter scheme} is used when $R_2<R_{\rm L}$. Then,

\begin{equation}
    \dot{M}_{\rm RLOF}= -\dot{M}_0 \exp\bigg( \frac{R_2-R_{\rm L}}{H_{\rm p}/\gamma(Q)} \bigg).
    \label{eq:MTR}
\end{equation}
\noindent Here $H_{\rm p}$ is the pressure scale height at the photosphere of the donor and

\begin{equation}
    \dot{M}_0= \frac{2\pi}{\exp(1/2)} F_1(Q) \frac{R_{\rm L}^3}{GM_2} \bigg(\frac{k_{\rm B}T_{\rm eff}}{m_{\rm p}\mu_{\rm ph}}\bigg)^{3/2} \rho_{\rm ph}
\end{equation}

\noindent is the MTR of a binary in which the donor star fills exactly its Roche Lobe. $M_2$ and $T_{\rm eff}$ are the mass and the effective temperature of the donor respectively, and $\mu_{\rm ph}$ and $\rho_{\rm ph}$ are the mean molecular weight and density at its photosphere. $m_{\rm p}$ is the proton mass, and $k_{\rm B}$ is the Boltzmann constant. The two fitting functions are

\begin{equation}
    F_1(Q)= 1.23 + 0.5 \log{Q}, \quad 0.5\lesssim Q \lesssim 10
\end{equation}

\noindent and

\begin{equation}
    \gamma(Q)=
    \begin{cases}
    0.954+0.025\log{Q}-0.038(\log{Q})^2 & 0.04\lesssim Q \lesssim 1 \\
    0.954+0.039\log{Q}-0.114(\log{Q})^2 & 1\lesssim Q \lesssim 20,
    \end{cases}
\end{equation}

\noindent where $Q= M_2/M_1$. 

The {\it Kolb-Ritter scheme} is used when $R_2>R_{\rm L}$. Then,

\begin{equation}
\begin{split}
    \dot{M}_{\rm RLOF}= &-\dot{M}_0 - 2\pi F_1(Q) \frac{R_L^3}{GM_2} \\
    &\times \int_{P_{\rm ph}}^{P_{\rm RL}} \Gamma_1^{1/2} \bigg( \frac{2}{\Gamma_1+1} \bigg)^{(\Gamma_1+1)/(2\Gamma_1-2)} \bigg(\frac{k_BT}{m_p\mu}\bigg)^{1/2} dP. \label{MTR_KR}
\end{split}
\end{equation}

\noindent Here $\Gamma_1$ is the first adiabatic exponent, and $P_{\rm ph}$ and $P_{\rm RL}$ are the pressure \referee{values} at the photosphere and at the $R_L$ \referee{value of the radius}, respectively.

\subsection{Magnetic braking} \label{subsec:Mbraking}

The orbital angular momentum loss due to MB is modeled after the prescriptions given by \citet{2019MNRAS.483.5595V} (hereafter Van19) and \citet{2019ApJ...886L..31V}. The prescriptions in Van19 are defined by the expression:

\begin{equation}
    \frac{dJ_{\rm MB}}{dt} = \frac{dJ_{\rm MB,Sk}}{dt} \bigg( \frac{\omega_2}{\omega_\odot} \bigg)^{\beta} \bigg( \frac{\tau_{\rm conv}}{\tau_{\rm \odot,conv}} \bigg)^{\xi} \bigg( \frac{\dot{M}_{\rm 2,wind}}{\dot{M}_{\rm \odot,wind}} \bigg)^{\alpha},
\label{MBs}
\end{equation}

\noindent where $dJ_{\rm MB,Sk}/dt$ is the MB law derived by \citet{1983ApJ...275..713R}:

\begin{equation}
    \frac{dJ_{\rm MB,Sk}}{dt} = -3.8 \times 10^{-30} M_2 R_{\odot}^4 \bigg( \frac{R_2}{R_{\odot}} \bigg) ^{\gamma_{\rm mb}} \omega_2^{3} ~~ \rm{dyn \; cm},
\end{equation}

\noindent $\gamma_{\rm MB}=4$ corresponds to the standard Skumanich law  \citep{1981A&A...100L...7V} and $\omega_2$ is the angular velocity of the donor star. In this work, we use $\omega_{\odot}=3 \times 10^{-6}~\rm{s^{-1}}$, $\tau_{\rm \odot,conv}= 1.537 \times 10^{6}~\rm{s}$\footnote{ We have found this value by detailed modeling of the Sun employing our stellar code. In this way, we handle the different MB laws consistently, which is of key relevance for our simulations. In the literature (e.g., Van19) it has been given a value of $\tau_{\rm \odot,conv}=2.8 \times 10^{6}~\rm{s}$. } and $\dot{M}_{\rm \odot,wind}=2.54 \times 10^{-14}~M_{\odot}~{\rm  yr}^{-1}$ \citep{2006ima..book.....C}.

According to Van19, $\beta$, $\xi$, and $\alpha$ in Eq.~(\ref{MBs}) can take different values, resulting in four MB laws with different strengths:

\begin{equation}
    (\beta, \xi, \alpha)=
    \begin{cases}
    (0,0,0) & \text{MB0 -- standard MB} \\
    (0,2,0) & \text{MB2 -- convection-boosted MB} \\
    (0,2,1) & \text{MB3 -- intermediate MB} \\
    (2,4,1) & \text{MB4 -- wind-boosted MB} 
    \end{cases} \label{abg}
\end{equation}

Short after presenting these MB prescriptions, \citet{2019ApJ...886L..31V} proposed another expression called Convection And Rotation Boosted (CARB~MB) that, according to them, better agrees with all well-observed systems:

\begin{equation}
    \begin{split}
    \frac{dJ_{\rm CARB~MB}}{dt} = &- \frac{2}{3} \dot{M}_{\rm 2,wind}^{-1/3} R_2^{14/3} (v_{\rm esc}^2 + 2\omega_2^2 R_2^2 / K_2^2)^{-2/3} \\
    &\times \omega_{\odot} B_{\odot}^{8/3} \bigg( \frac{\omega_2}{\omega_\odot} \bigg)^{11/3} \bigg( \frac{\tau_{\rm conv}}{\tau_{\odot,{\rm conv}}} \bigg)^{8/3},
    \end{split} 
\label{CARB-MB}
\end{equation}

\noindent where $v_{\rm esc}^2= 2GM_{2}/R_{2}$ is the surface escape velocity, and $B_\odot= 1~G$. $K_2= 0.07$ is a constant obtained from a grid of simulations by \citet{2015ApJ...798..116R} that sets the limit where the rotation rate begins to play a significant role. 

\referee{Let us remark that, for the MB to operate on the donor star it is necessary the presence of an outer convective zone (OCZ). In our code, we assume that MB fully acts if $M_{\rm OCZ}/M_{2}>10^{-2}$, where $M_{\rm OCZ}$ is the mass of the OCZ  \citep{2002ApJ...565.1107P}.}

For the eddy turnover timescale, we use the expression given in \citet{2002MNRAS.329..897H}:

\begin{equation}
    \tau_{\rm conv} = 0.2989 \bigg[ 
    \bigg(\frac{M_{\rm env}}{M_{\odot}}\bigg)
    \bigg(\frac{R_{\rm env}(R_2-R_{\rm env}/2)}{R_{\odot}^2}\bigg)
    \bigg(\frac{L_{\odot}}{L_2}\bigg) \bigg]^{1/3} {\rm yr}, \label{tauconvec}
\end{equation}

\noindent where $R_{\rm env}$ and $M_{\rm env}$ are the depth and mass of the convective envelope of the donor star respectively, and  \referee{$L_2$ is its luminosity}.

\subsection{Tides} \label{subsec:tides}

We included the effects of tides to relax the standard instantaneous synchronisation assumption. To do so, we proceed as follows. In the first model of a \referee{simulation}, we assume the donor is synchronised with the orbit. From then on, we introduce MB acting on the rotation of the donor, \referee{which is coupled to the orbit of the system by tides}. This spin-orbit coupling is considered only during detached states; on the contrary, if the system is undergoing \referee{a} RLOF episode, we assume instantaneous synchronisation. 

We consider the spin-orbit coupling following \citet{2014MNRAS.444..542R}. In this work, we assume circular orbits and coplanarity; accordingly, the set of \referee{differential equations we solve} reduces to

\begin{equation}
\label{eq:Omega_orbita}
 \frac{da}{dt}= -6 \bigg(\frac{K}{T}\bigg) q(1+q) \bigg(\frac{R_{2}}{a}\bigg)^8 a
 \bigg[1 - \frac{\omega_2}{\Omega} \bigg]+\bigg(\frac{da}{dt}\bigg)_{gw},
\end{equation}

\begin{equation}
\label{eq:rotacion}
 \frac{d\omega_2}{dt}= 3 \bigg( \frac{K}{T} \bigg)
 \frac{q^2}{k^2} \bigg( \frac{R_{2}}{a} \bigg)^6
 \Omega \bigg[1 - \frac{\omega_2}{\Omega} \bigg] + \bigg(\frac{d\omega_2}{dt}\bigg)_{\rm mb},
\end{equation}

\noindent where $a$ and $\Omega$ are the semi-major axes and the mean angular velocity of the orbit respectively, $q= 1/Q$, and $k$ is the radius of gyration of the donor star, which describes its moment of inertia $I$ as $I= k^{2} M_{2} R_{2}^{2}$. $(da/dt)_{gw}$ is due to gravitational wave radiation. $\big(K/T\big)$ is the tidal timescale, which strongly depends on the structure of the star. We use different prescriptions during the evolution of the donor, following the expressions given in \citet{2002MNRAS.329..897H}:

\begin{equation}
     \bigg( \frac{K}{T} \bigg)=
    \begin{cases}
     \frac{2}{21} \frac{F_{\rm conv}}{\tau_{\rm conv}} 
 \frac{M_{\rm env}}{M_{2}}~{\rm yr}^{-1} & \text{convective envelope} \\
    1.9782 \times 10^4 \frac{M_2 R_2^2}{a^5} (1+q)^{5/6} E_2~{\rm yr}^{-1} & \text{radiative envelope} \\
    2.564 \times 10^{-8} k^2 \big( \frac{L_2}{M_2} \big)^{5/7}~{\rm yr}^{-1} & \text{white dwarf} 
    \end{cases}
    \label{eq:KT}
\end{equation}

\noindent $F_{\rm conv}$ is the fraction of the convective cells which contribute to the damping: $F_{\rm conv}= {\rm min} \big[ 1, \big( \frac{P_{\rm tid}}{2\tau_{\rm conv}} \big) \big]$, $\frac{1}{P_{\rm tid}}= \big| \frac{1}{P_{\rm orb}} - \frac{1}{P_{\rm spin}}\big| $ and $E_2= 1.592\times 10^{-9} M_2^{2.84}$ \citep{1975A&A....41..329Z}.

The last term in Eq.~{\eqref{eq:rotacion}} changes according to the considered MB prescription (MB0, MB2, MB3, or CARB~MB).%{\sout{\referee{stands for the coupling of MB with the rotation of the donor.}}} 

\subsection{Stellar wind}

For modeling the stellar wind, we use the expression from \citet{1975psae.book..229R}:

\begin{equation}
    \dot{M}_{\rm 2,wind}= 4 \times 10^{-13} ~\eta ~\bigg( \frac{R_{2}}{R_{\odot}} \bigg) \bigg( \frac{L_{2}}{L_{\odot}} \bigg) \bigg( \frac{M_{\odot}}{M_{2}} \bigg)  M_{\odot} \; {\rm yr}^{-1},
    \label{eq:vientos}
\end{equation}

\noindent where $\eta=1$ is the efficiency of wind loss.

\subsection{\referee{Binary models}}\label{sec:binarymod}

\referee{For all the models presented in this paper,} we consider donor and neutron stars of $1.25 M_{\odot}$ and $1.3 M_{\odot}$ respectively. We use a metallicity value $Z=0.0139$ (\citealt{2021A&A...653A.141A}) and have carefully tuned the hydrogen abundance and the mixing length parameter to reproduce the present Sun. This results in $\alpha_{\sc mlt} = l/H_{\rm p} = 1.55$, where $l$ is the mixing length, and a hydrogen abundance \referee{of} $X=0.7238$. We start our simulations assuming the donor star is on the Zero Age Main of Sequence (ZAMS). We neglect overshooting. The orbital evolution is calculated assuming $\alpha=1$ and $\beta=1$. \referee{Therefore,} if the donor star transfers mass at a rate lower than $\dot{M}_{\rm Edd}$, the system evolves in {\it conservative} conditions.

The sequences are computed up to an age of 14000~M{\rm yr}, or until the donor star reaches a very low luminosity ($L_2 <10^{-7}~L_{\odot}$), or a very low mass ($M_2 <0.015~M_{\odot}$). We also stop the calculations if the MTR reaches the value of $10^{-2}~M_{\odot}~{\rm yr}^{-1}$. \referee{Higher rates cannot be considered under the physics included in our code. Very high MTR corresponds to the regime described by Eq.~(\ref{MTR_KR}). \citet{1990A&A...236..385K} considered this treatment as plausible for a donor star that overfills its Roche lobe up to $(R_2-R_{\rm L})/H_{\rm p} \approx 10$, which in our models provide MTR values of the order of the limiting rate quoted above. It is expected that even stronger MTRs lead to the formation of a common envelope. This process is beyond the scope of this paper.}

We constructed sequences under the Skumanich (MB0), Convection-boosted (MB2), Intermediate (MB3), and Convection And Rotation-Boosted (CARB~MB) prescriptions. The initial orbital period $P_{\rm orb,i}$ extends from 0.5 to 328.42~d for MB3; from 0.5 to 738.94~d for MB0; and from 0.5 to 1108.41~d for MB2 and CARB~MB. \referee{In the main set of models}, the initial orbital period step is 0.176 in $\log_{10}(P_{\rm orb,i})$. \referee{To study the formation of UCXBs, we refine the grid of initial orbital periods in the range that leads to this kind of system (see Sec.~\ref{sec:ucxbs}). }

%--------------------------------------------------------------------------------
\section{Results} \label{sec:results}

\begin{figure*}
\centering  
\begin{subfigure}[b]{0.475\textwidth}
    \includegraphics[width=\textwidth]{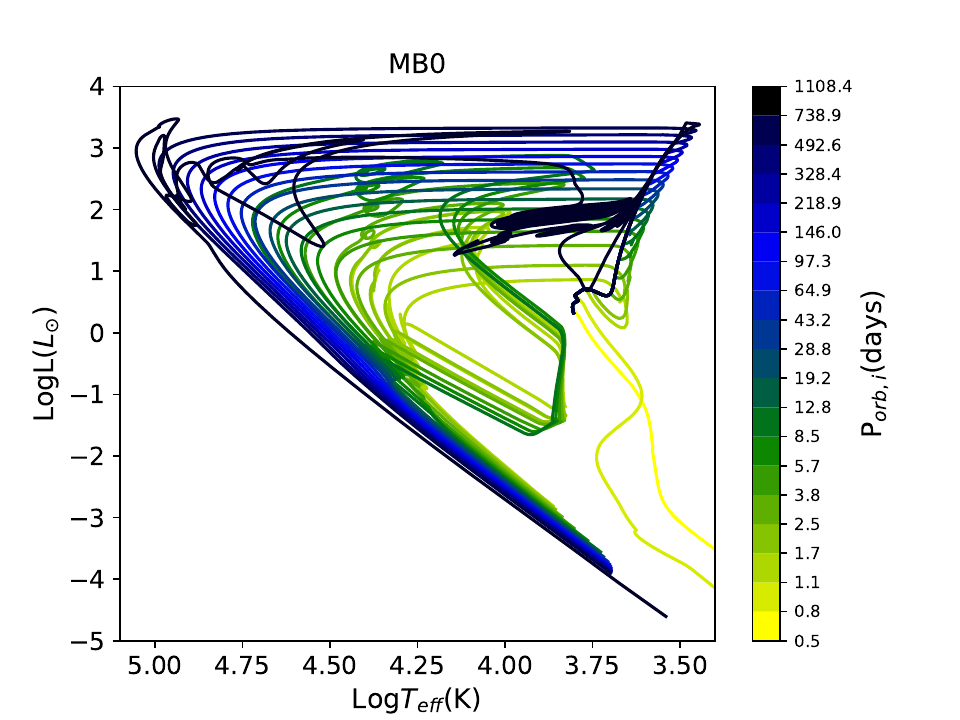}
    \caption{}
    \label{HR_MB0}
\end{subfigure}
\begin{subfigure}[b]{0.475\textwidth}
    \includegraphics[width=\textwidth]{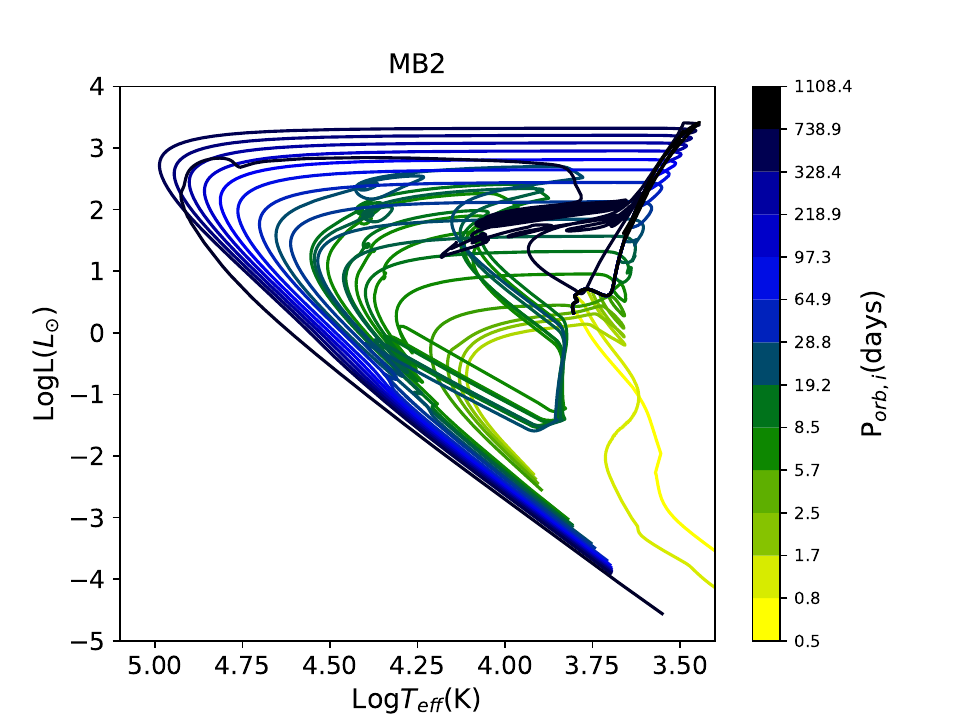}
    \caption{}
    \label{HR_MB2}
\end{subfigure}
\begin{subfigure}[b]{0.475\textwidth}
    \includegraphics[width=\textwidth]{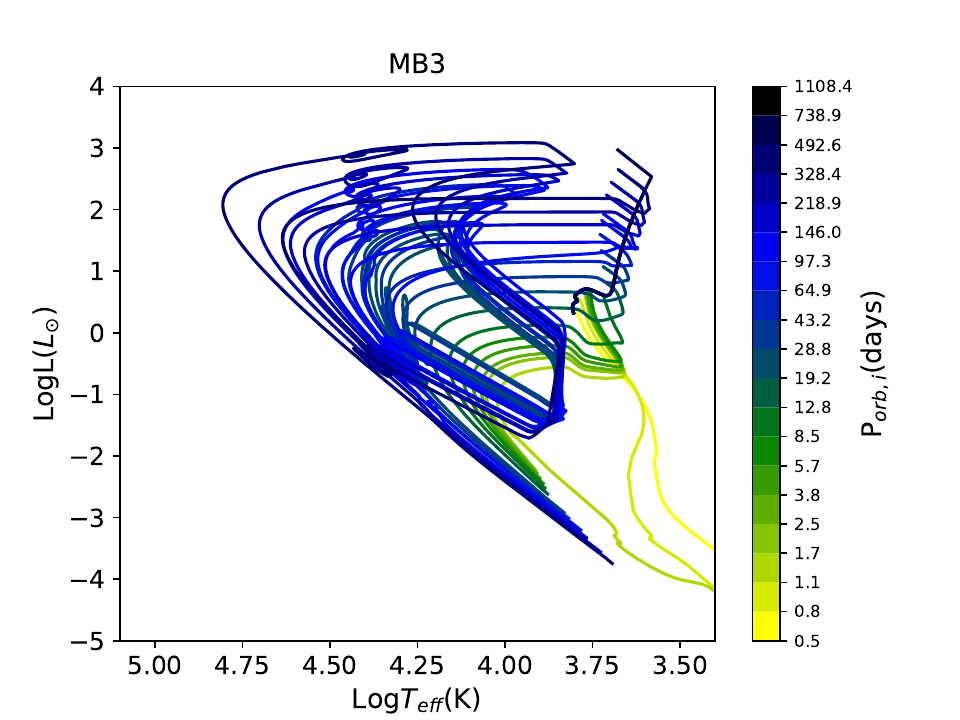}
    \caption{}
    \label{HR_MB3}
\end{subfigure}
\begin{subfigure}[b]{0.475\textwidth}
    \includegraphics[width=\textwidth]{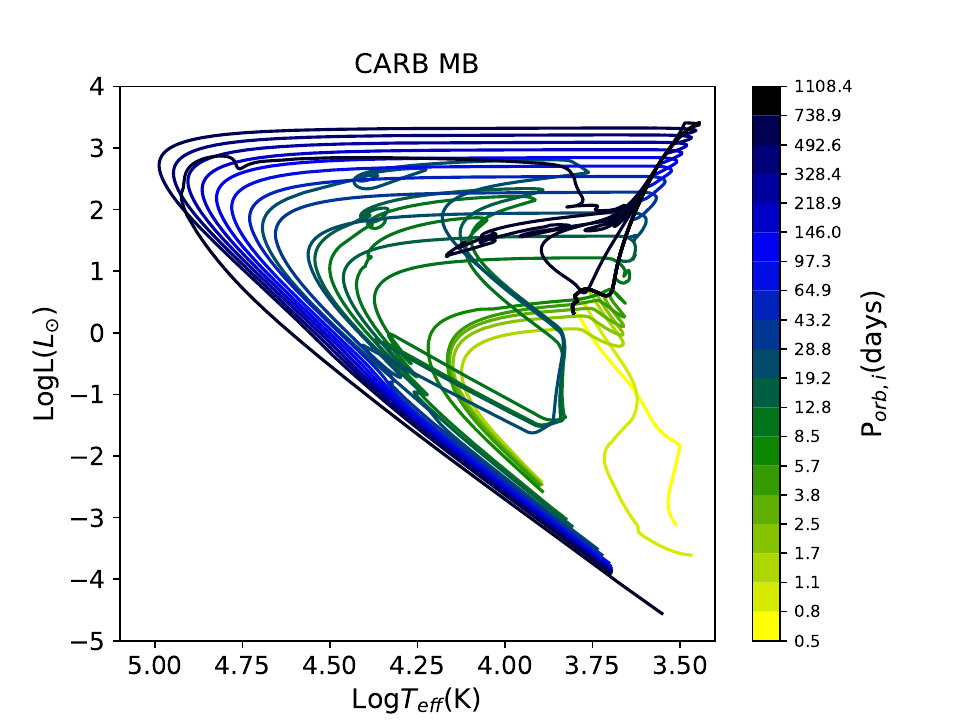}
    \caption{}
    \label{HR_MB4}
\end{subfigure}
\caption{Evolutionary tracks of binary systems containing a $1.25~M_{\odot}$ donor star together with a $1.3~M_{\odot}$ NS. Colors from yellow to black indicate the initial orbital period, going from the lowest to the largest value. Each panel represents the evolution of a set of models under a particular MB prescription.}
\label{HR}
\end{figure*}

\referee{To explore the differences and similarities in the evolution of the systems under different MB prescriptions, in Fig.~\ref{HR} we show the evolutionary tracks of the donor stars in the Hertzsprung-Russell diagram (HRD) corresponding to the main set of calculations (see Sect. \ref{sec:binarymod}). Each panel shows the evolution under one of the considered MB prescriptions. The color bar indicates the initial orbital period value, $P_{\rm orb,i}$, in the simulations. At first glance, we note that the sequences obtained with MB3 are different from those computed with the other three prescriptions. One of the main reasons for the occurrence of this disparity is that MB3 induces stronger angular momentum losses, resulting in higher MTR values (see below) which, for $P_{\rm orb,i}>$328~d exceed the maximum allowed MTR value (see Sect.~\ref{sec:binarymod}). On the contrary, for the other three MB prescriptions, it is possible to evolve systems with $P_{\rm orb,i}$ considerably larger.}

As a consequence of the high MTR suffered by donors evolved under MB3, for a given value of $P_{\rm orb,i}$, the evolution leads to the formation of WDs lighter than those obtained under the other prescriptions. Another remarkable characteristic of the models computed with the MB3 is the evolution at an almost constant radius at the onset of the mass transfer episode (see the right upper corner in panel (c), Fig.~\ref{HR}). This is due to the high value of the mass transfer rate on a very short timescale ($M/\dot{M}\approx 100$~y, see Fig.~\ref{mtr}, panel c). This scale is too short for the star to change its radius, while its effective temperature and luminosity change appreciably. 

\referee{For systems with the largest $P_{\rm orb,i}$ evolved under MB0, MB2, and CARB~MB, it has been possible for the donor star to grow its helium core mass above approximately $0.45\; M_{\odot}$, thereby enabling the occurrence of a helium flash, leading to the formation of a CO WD (depicted by the black tracks in Fig.~\ref{HR}). Note that the cooling track corresponding to these donors separates from the others. This behavior arises from the significant discrepancy in the chemical composition, as donors with shorter initial orbital periods end their evolution as HE WDs. On the contrary, in the case of MB3, none of the donors experienced a helium flash, resulting in the formation of HE WDs in all cases.}

\begin{figure*}
\centering
\begin{subfigure}[b]{0.475\textwidth}
    \includegraphics[width=\textwidth]{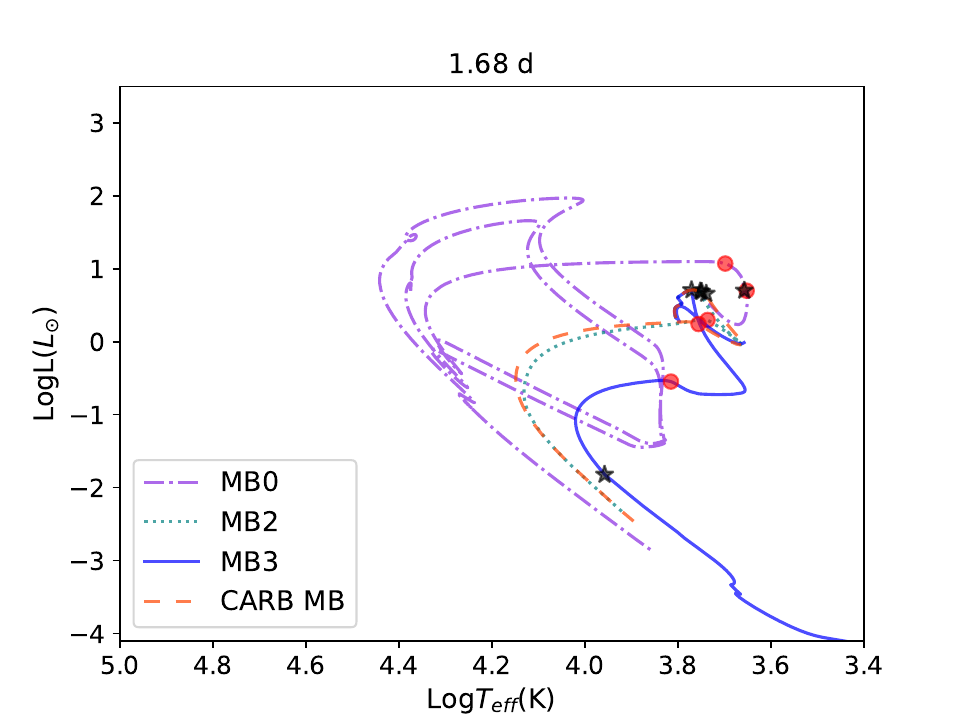}
\end{subfigure}
\begin{subfigure}[b]{0.475\textwidth}
    \includegraphics[width=\textwidth]{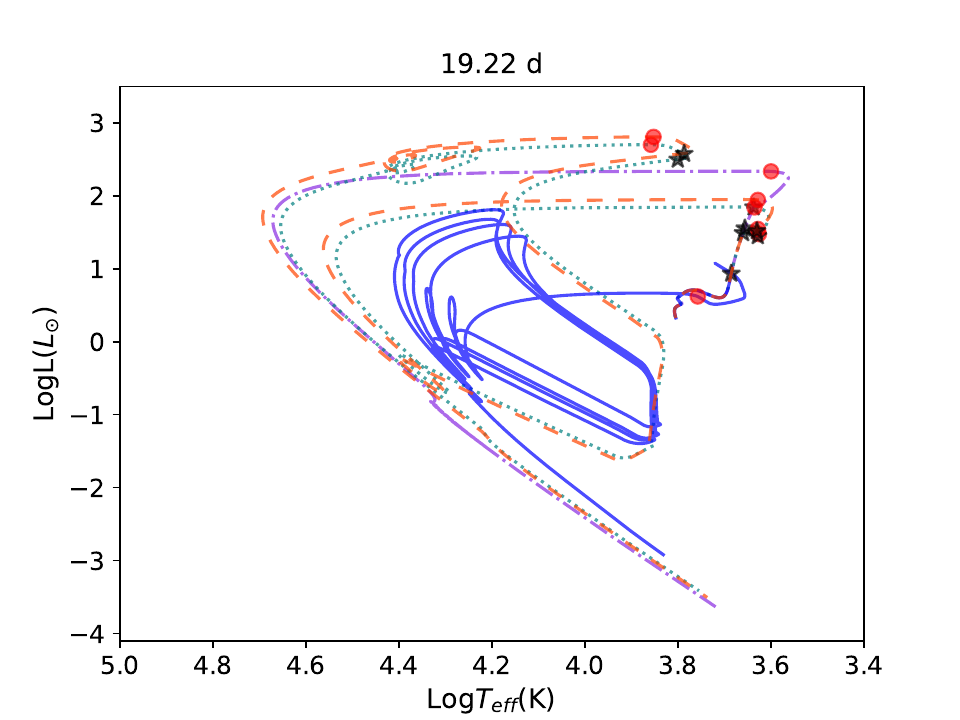}
\end{subfigure}
\begin{subfigure}[b]{0.475\textwidth}
    \includegraphics[width=\textwidth]{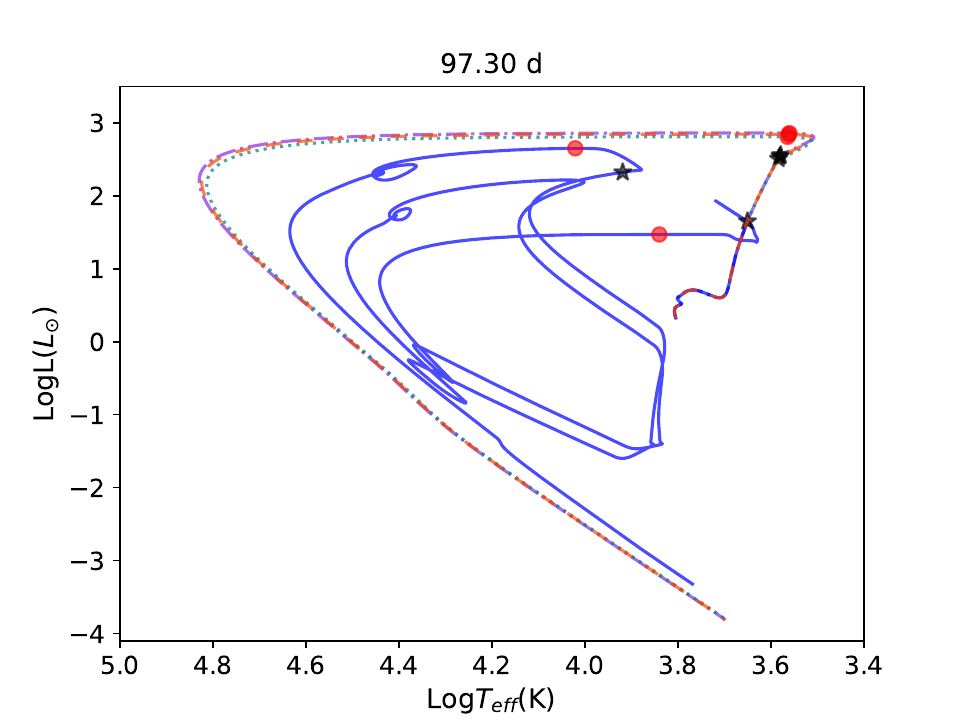}
\end{subfigure}
\begin{subfigure}[b]{0.475\textwidth}
    \includegraphics[width=\textwidth]{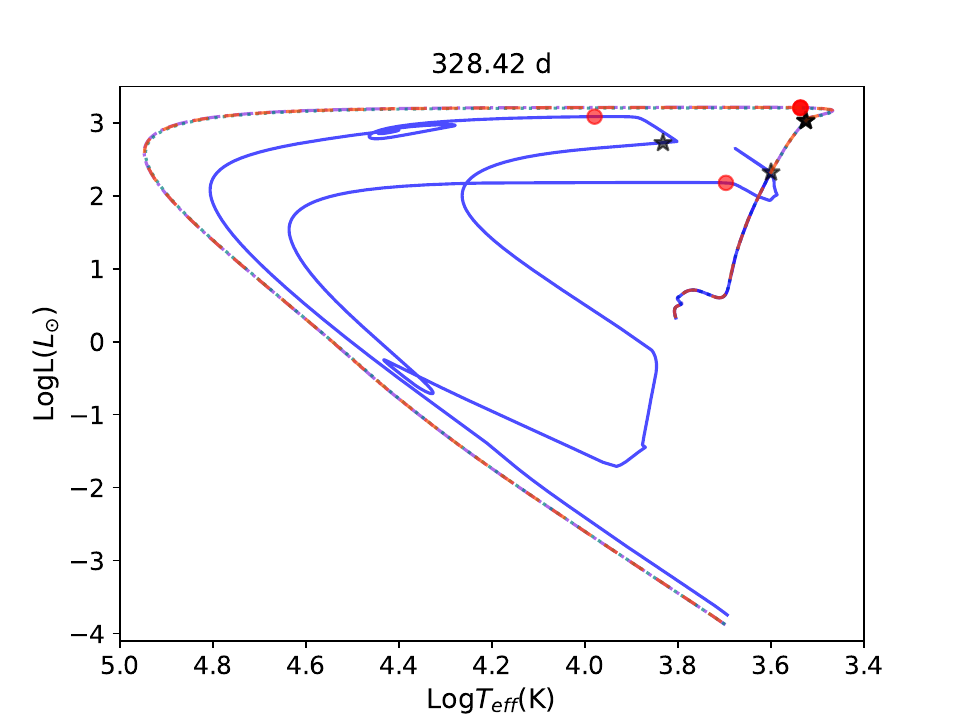}
\end{subfigure}
\caption{Hertzsprung-Russell diagram corresponding to the donor stars of binaries with $P_{\rm orb,i}$= 1.68, 19.22, 97.30 and 328.42~d. Black stars indicate the start of mass transfer, and red circles its end. Each line style represents a different MB prescription. Notice that, for the cases of $P_{\rm orb,i}$= 97.30 and 328.42~d, the tracks corresponding to MB0, MB2, and CARB~MB are almost indistinguishable from each other.}
\label{HR_4}
\end{figure*}

\referee{Let us explore the effects of changing the MB prescriptions on the global evolution of these binaries. For this purpose, we select sequences corresponding to initial orbital period values of 1.68, 19.22, 97.30, and 328.42~d.
Fig.~\ref{HR_4} shows the HRD of the selected donors. Each panel corresponds to a given $P_{\rm orb,i}$ value and has four tracks, plotted with a different line style and color that indicates a specific MB prescription. The black stars indicate the start of mass transfer and the red circles its end. As can be noticed from this figure, for the shortest selected initial orbital period, there is a remarkable disparity in the results obtained with different MB prescriptions. Specifically, for models with MB0, the donor evolves undergoing several thermonuclear flashes, ending as a HE WD on a $P_{\rm orb} \sim 10$~d orbit. Donors under MB2 and CARB do not suffer flashes and end with a $P_{\rm orb} \sim 1$~d orbit. Remarkably, the same model under MB3 ends as a UCXB. On the contrary, as $P_{\rm orb,i}$ increases, systems evolved under MB0, MB2, and CARB~MB become more similar to each other, while those evolved under MB3 remain different. }

\referee{Let us discuss the physical reason for this behavior by analyzing Eqs.~(\ref{MBs}), (\ref{abg}), and (\ref{CARB-MB}). MB2 differs from MB0 in that the former considers the factor $(\tau_{\rm conv}/\tau_{\rm \odot,conv})^2$. MB3, on the other hand, incorporates the same factor as MB2 but also includes a dependence on the stellar wind mass loss $(\dot{M}_{\rm 2,wind}/\dot{M}_{\rm \odot,wind})$. Finally, CARB MB has a more complicated dependence on these factors, and because of this reason, it will be analyzed separately.}

\begin{figure*} 
\hspace*{-0.85in}
\centering
 \includegraphics[scale=0.43]{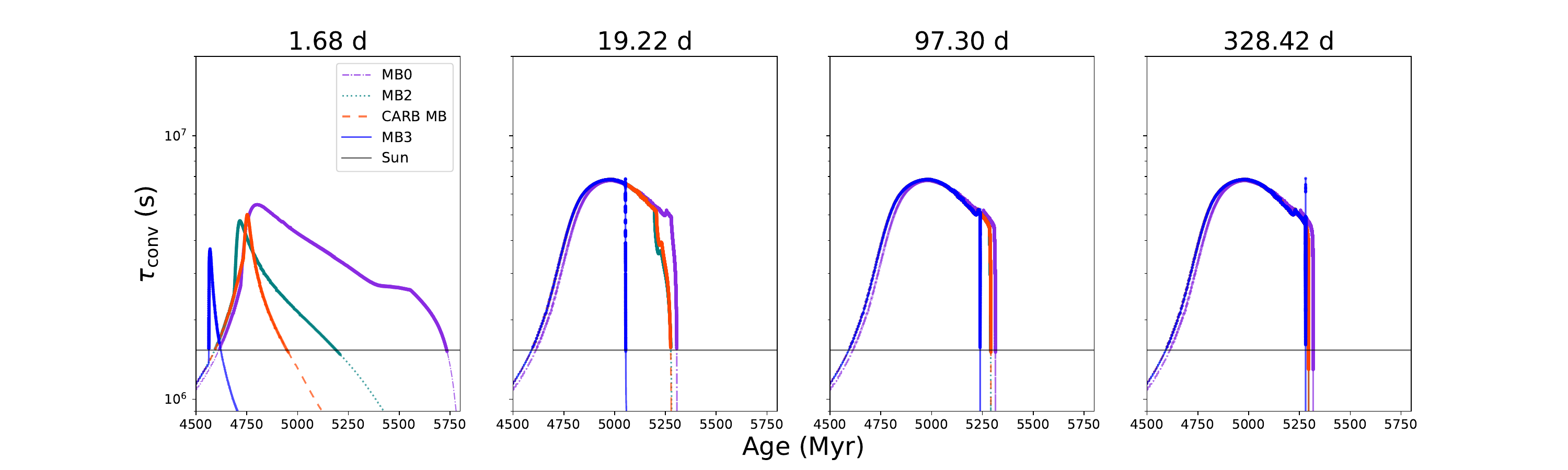}
\caption{Thermal timescale evolution of the systems in Fig.~~\ref{HR_4}. We restrict the horizontal axis to the age in which MB operates with a significant strength. Thick lines correspond to stages in which $M_{\rm OCZ}/M_{2}>0.01$. The thin continuation of these lines stands for models with thinner OCZs for which MB is not relevant. The solid horizontal line corresponds to the thermal timescale of the Sun.}
\label{tauconvs}
\end{figure*}

\referee{Given the dependence of the strength of MBs with $\tau_{\rm conv}$, we studied the evolution of this quantity using Eq.~(\ref{tauconvec}). In Fig.~\ref{tauconvs} we present the $\tau_{\rm conv}$ corresponding to the same models employed in Fig.~\ref{HR_4}. The time interval considered in these plots corresponds to the stages in which most of the mass is transferred and simultaneously, the OCZ of the donor star is wide enough to allow MB to act. As can be seen, for the case of the shortest selected orbital period, $\tau_{\rm conv}$ exhibits different behavior for each MB prescription, directly influencing the strength of the angular momentum losses.} %{\color{cyan} This difference leads to a disparity between models under MB2 and MB0} Y LA CARB?} \ale{(en la CARB tambien, por eso la analizamos aparte. Dejaria la verson anterior)}.
\referee{As the initial orbital period increases, the $\tau_{\rm conv}$ values corresponding to the models with MB0, MB2 and CARB~MB become more similar to each other, becoming indistinguishable for the models with the longest selected $P_{\rm orb,i}$. For the case of MB3, $\tau_{\rm conv}$ differentiates from the other prescriptions at intermediate values of $P_{\rm orb, i}$, but exhibits a more similar behavior at larger $P_{\rm orb,i}$. In all cases, the corresponding MB3 models undergo stronger braking due to the effect of the stellar wind as the radius and luminosity of the donor star increase. This effect is more prominent in the cases of long $P_{\rm orb,i}$, where the donor is climbing the red giant branch. The wind, and then MB3, is more intense, implying greater angular momentum losses than the other MB prescriptions, leading to an earlier RLOF (see Eqs.~\ref{MBs}, and \ref{abg} and Fig.~\ref{HR_4}). Therefore, systems under MB3 deeply differentiate from the rest of the prescriptions at all the considered initial orbital periods, even though $\tau_{\rm conv}$ becomes similar under all prescriptions at large $P_{\rm orb,i}$}.

\begin{figure*} 
\hspace*{-0.85in}
\centering
 \includegraphics[scale=0.43]{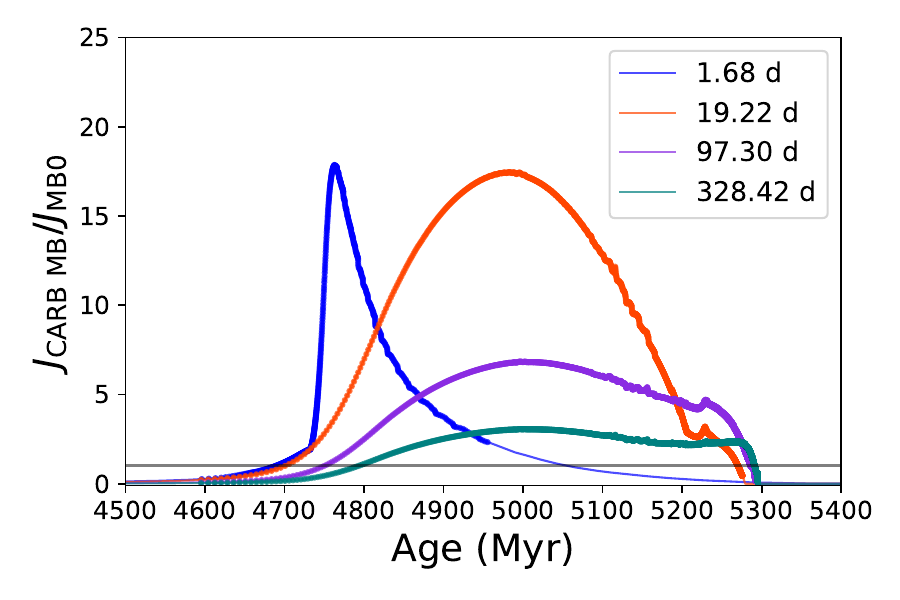}
\caption{Ratio of the strength of $J_{\rm CARB \; MB}$ to $J_{\rm MB0}$ as a function of time for the models included in Fig.~~\ref{HR_4}. The thick portion of the lines corresponds to stages in which $M_{\rm OCZ}/M_{2}>0.01$. The horizontal line signalizes $J_{\rm CARB\; MB} / J_{\rm MB0} = 1$.}
\label{strenght_carb_mb0}
\end{figure*}

\begin{figure*} 
\centering
\begin{subfigure}[b]{0.475\textwidth}
    \includegraphics[width=\textwidth]{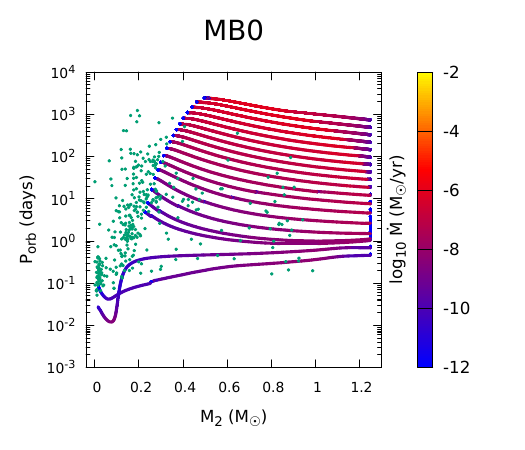}
    \caption{}
    \label{MP_MB0}
\end{subfigure}
\begin{subfigure}[b]{0.475\textwidth}
    \includegraphics[width=\textwidth]{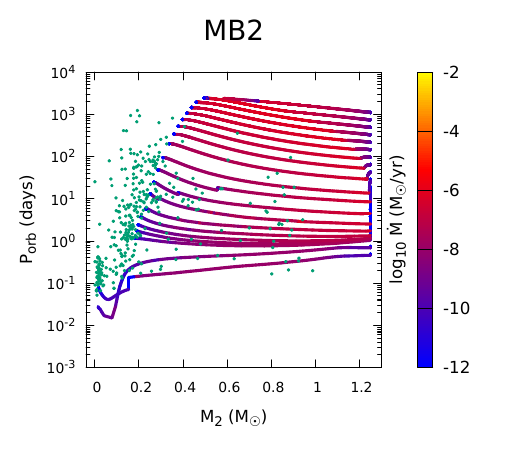}
    \caption{}
    \label{MP_MB2}
\end{subfigure}
\begin{subfigure}[b]{0.475\textwidth}
    \includegraphics[width=\textwidth]{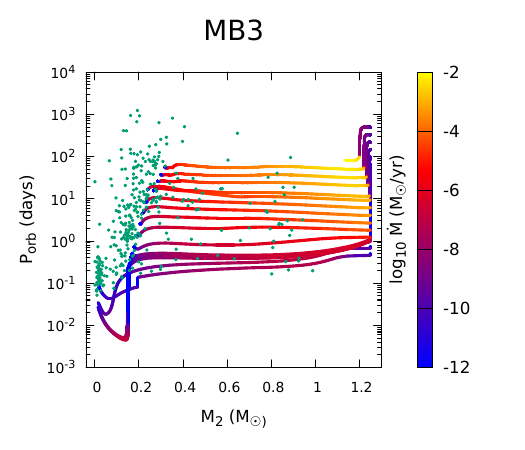}
    \caption{}
    \label{MP_MB3}
\end{subfigure}
\begin{subfigure}[b]{0.475\textwidth}
    \includegraphics[width=\textwidth]{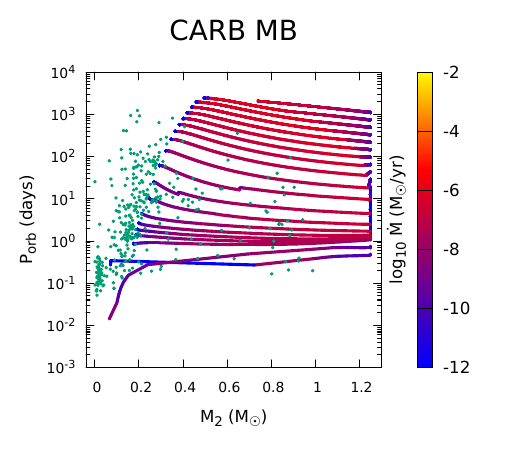}
    \caption{}
    \label{MP_MB4}
\end{subfigure}
\caption{Orbital period as a function of the donor star mass. The color bar represents the MTR. Each panel corresponds to a given MB prescription. Green dots correspond to the median mass of pulsar companions taken from the ATNF Pulsar Catalogue.}
\label{mtr}
\end{figure*}

\referee{The CARB MB prescription has a more complicated expression (see Eq.~\ref{CARB-MB}). 
To compare it with the standard treatment provided by MB0, we present the ratio of the strengths of CARB MB to MB0 (Fig.~\ref{strenght_carb_mb0}) corresponding to the same models employed in Fig.~\ref{HR_4}. 
We show only the portion of the evolution in which the OCZ is thick enough to allow for the full action of MB on the donor star rotation (when $M_{\rm OCZ}/M_{2}>0.02$). From the inspection of Fig.~\ref{strenght_carb_mb0}, it is clear that at short initial orbital periods, the difference between CARB MB and MB0 is pronounced, and it becomes noticeably smaller for long $P_{\rm orb,i}$ values.}

\referee{Let us now discuss the reason for the similarity of the tracks computed with MB0, MB2, and CARB MB in the case of long initial orbital periods. The orbital angular momentum of these systems is proportional to $a^{1/2}$ or, equivalently, to $P_{\rm orb}^{1/3}$. Magnetic braking acts on the donor star and has no direct dependence on the orbital period, but couples to the orbital motion by tidal interactions (see Eq.~\ref{eq:rotacion}). Therefore, the longer the orbital period, the less significant the angular momentum loss due to MB. Thus, the differences in the strength of MB0, MB2, and CARB MB are proportionally less important. This naturally leads to the occurrence of very similar tracks for these three prescriptions. In the case of MB3, the dependence on the stellar wind is significant enough to cause the models under this prescription to differentiate from the other MBs, even at large orbital periods.}

%\vspace{0.5cm}

%\vspace{0.5cm}

%\begin{figure*} 
%\hspace*{-0.85in}
%\centering
% \includegraphics[scale=0.43]{TCONV.pdf}
%\caption{ ELIJAN CUAL QUEDA MEJOR }
%\label{tauconvs}
%\end{figure*}

\begin{table}
   \centering
   \caption{NS final mass ($M_\odot$) considering different MB prescriptions and initial orbital periods.}
   \title{NS}
\begin{tabular}{|c ||c | c |c |c | c | c |c | c | c |}
    \hline
    \parbox[t]{3mm}
      \bf $P_{\rm orb,i}$ & \bf MB0 & \bf MB2 & \bf MB3 & \bf CARB~MB \\[1 ex] \hline
      \bf 1.68 & 2.086 & 1.984 & 1.682 & 1.905 \\ [1ex] \hline
     \bf 19.22 & 1.614 & 1.686 & 1.329 & 1.724\\ [1ex] \hline
    \bf 97.30 &  1.364 & 1.371 & 1.305 & 1.367 \\ [1ex] \hline
     \bf 328.42 & 1.319 & 1.320 & 1.301 & 1.319 \\ [1ex] \hline
     \hline 
\end{tabular}
\label{tabla}
\end{table}

\referee{In Fig.~\ref{mtr}, we present the results of the simulations depicted in Fig.~\ref{HR}, now in the donor star mass versus orbital period plane, together with the masses of pulsar companions taken from the ATNF Pulsar Catalogue\footnote{\url{http://www.atnf.csiro.au/research/pulsar/psrcat}} (\citealt{2005AJ....129.1993M}). The color bar denotes the logarithm of the MTR in the simulations. As we can appreciate from this figure, as $P_{\rm orb,i}$ increases, so does the MTR. This effect is a direct consequence of the structure of the donor at the onset of the RLOF episode. The larger the value of $P_{\rm orb,i}$, the more evolved the donor star and the shorter the evolutionary timescale, leading to a larger MTR.}

\referee{It is evident from Fig. \ref{mtr} that MB3 yields MTR values at least one order of magnitude higher than the other MB prescriptions. Consequently, we obtain lighter donors. On the other hand, these MTR values greatly exceed the value that the NS can accrete (the Eddington critical rate, see Sec.\ref{sec:code}), resulting in less massive NSs compared to the other MB prescriptions, as detailed in Table~\ref{tabla}.}

%Furthermore, all systems under MB3 end up with $P_{\rm orb} < P_{\rm orb, i}$, allowing to reproduce binaries with final orbital periods up to $\simeq$55~d \ale{ESTO NO JUSTIFICA QUE LLEGUES HASTA ESTE VALOR DE Porb, SINO LA ALTA MTR, A LA QUE YA NOS REFERIMOS. NO PODEMOS CALCULARLOS XQ VAN A ENV COMUN, YA LO DIJIMOS}. Therefore, it is not possible to reach orbital periods as large as with the other prescriptions. Notably, MB0, MB2, and CARB~MB exhibit rather similar MTR values and allow to siumate systems with final orbital periods up to $\simeq$8~yr. \ale{LO MISMO...}}

\referee{Examining Fig.~\ref{mtr}, it can be seen that there is a region around $P_{\rm orb} \simeq $~1~d, occupied by many observed systems which in the case of MB0 models is reachable only by fine-tuning in $P_{\rm orb, i}$. Then, MB0 models predict a much scarcer population in contradiction with observations. This problem is partially mitigated with MB2, even more reduced with CARB~MB, and disappears with MB3. However, models with MB3 do not reach final periods larger than 55~d, while many observed systems are located well above this value. Models evolved with the other three prescriptions do not present this difficulty. Although this may be interpreted as a serious drawback for MB3, we remind the reader that in this work we study models with only one initial donor mass value. For more massive donors, the OCZ should be thinner and MB less effective. Therefore, these donors should be able to fill this orbital period region unreached by the models presented in this paper.}

\referee{As depicted in Fig.~\ref{mtr}, there are several MSPs located within a region where the simulations undergo mass transfer. A promising phenomenon for modeling these systems could be irradiation feedback (\citealt{1993A&A...277...81H}, \citealt{2004A&A...423..281B}). Irradiation feedback occurs when the donor star transfers mass onto the NS. The transferred matter emits X-ray radiation that illuminates the donor. If this star has an outer convective zone, its structure is unable to sustain the RLOF and the system becomes detached. Then, nuclear evolution may lead the donor to experience an RLOF again, leading to multiple mass transfer pulses in the system (see, e.g., \citealt{2014ApJ...786L...7B}). Irradiated systems can be detected as X-ray sources during RLOF states, or as MSPs once the mass transfer ceases. Including irradiation feedback in the models studied in the present paper should be relevant.}

\subsection{Formation of UCXBs}\label{sec:ucxbs}

\begin{figure*} 
\centering  
\begin{subfigure}[b]{0.475\textwidth}
    \includegraphics[width=\textwidth]{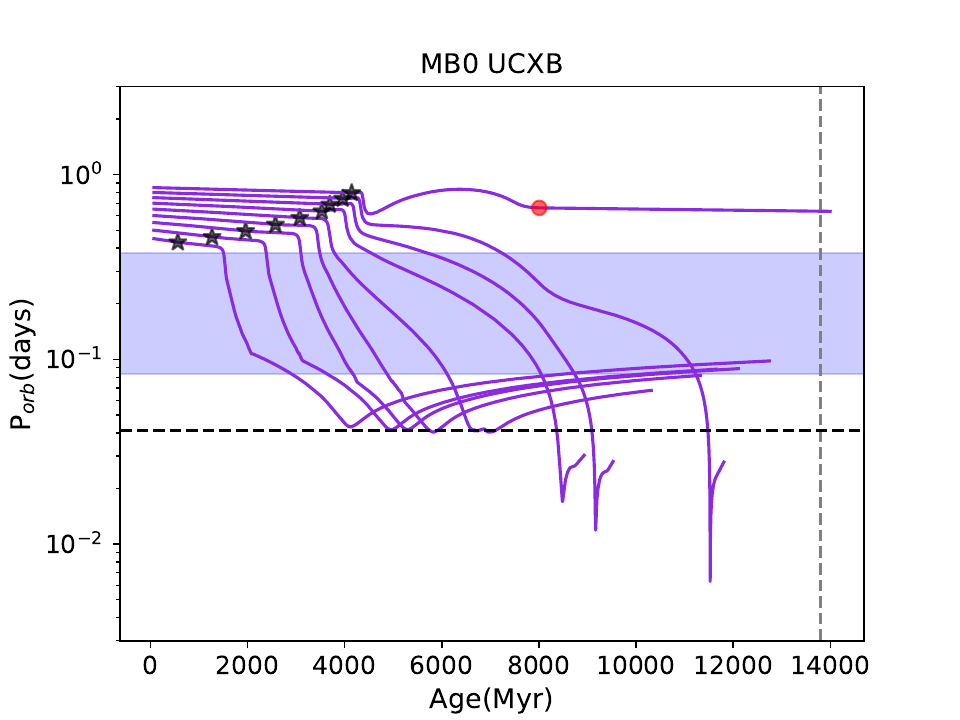}
    \caption{$P_{\rm orb,i}$ from 0.45 to 0.85~d}
    \label{UCXB_MB0}
\end{subfigure}
\begin{subfigure}[b]{0.475\textwidth}
    \includegraphics[width=\textwidth]{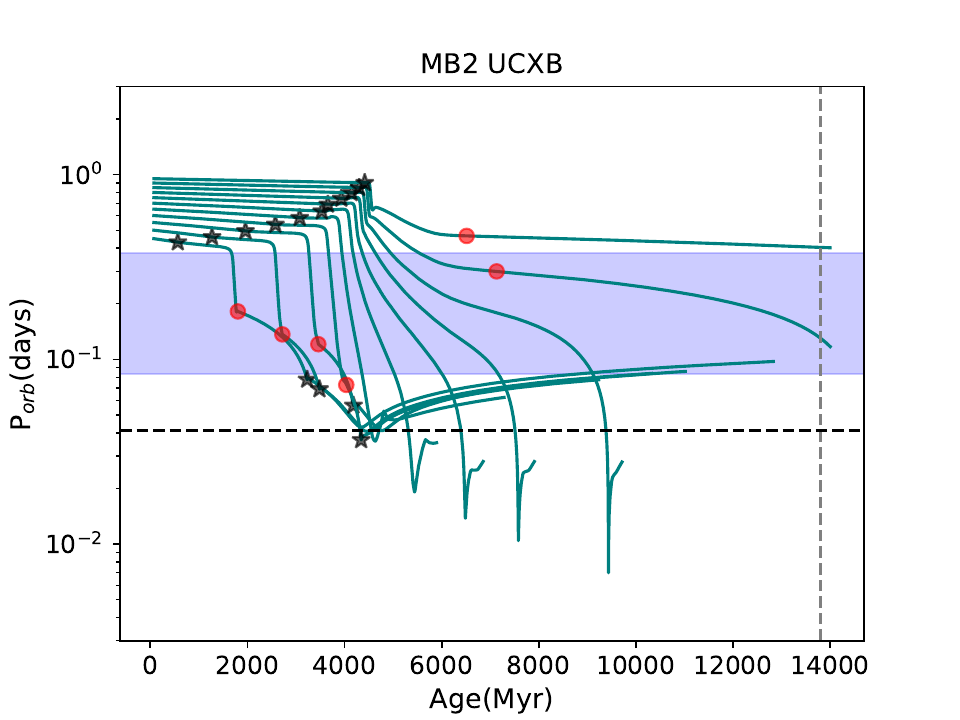}
    \caption{$P_{\rm orb,i}$ from 0.45 to 0.95~d}
    \label{UCXB_MB2}
\end{subfigure}
\begin{subfigure}[b]{0.475\textwidth}
    \includegraphics[width=\textwidth]{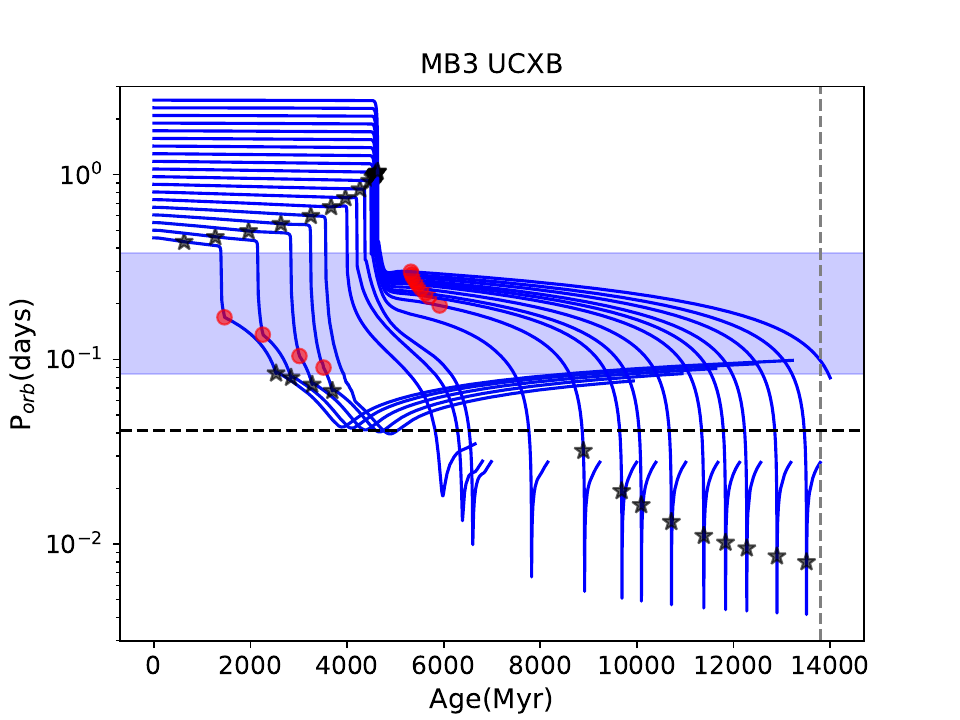}
    \caption{$P_{\rm orb,i}$ from 0.45 to 2.52~d}
    \label{UCXB_MB3}
\end{subfigure}
\begin{subfigure}[b]{0.475\textwidth}
    \includegraphics[width=\textwidth]{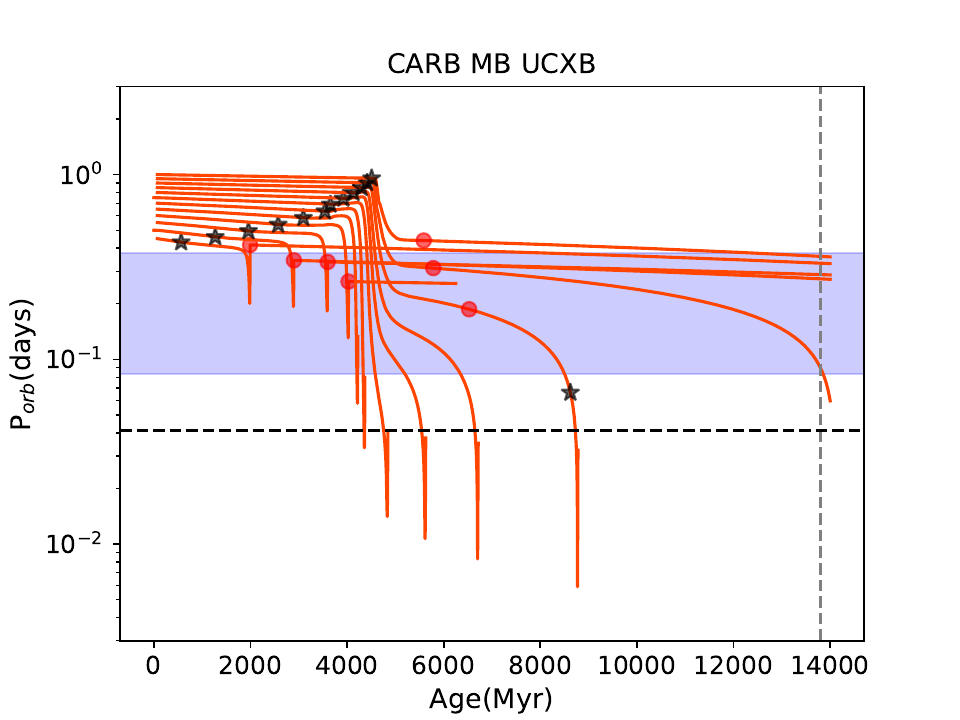}
    \caption{$P_{\rm orb,i}$ from 0.45 to 1.0~d}
    \label{UCXB_MB4}
\end{subfigure}
\caption{Temporal evolution of the orbital period of selected systems. Black stars indicate the start of mass transfer, and red circles its end. The horizontal light-blue band limits the region of orbital periods between 2 and 9~h. The horizontal black dashed line corresponds to 1~h. The vertical grey dashed line represents the age of the Universe.}
\label{ucxb}
\end{figure*}

\referee{As mentioned in Section~\ref{intro}, the formation of UCXBs under the MB0 prescription requires fine-tuning of the initial orbital period. Here, we aim to assess the capacity of MB2, MB3, and CARB MB models in mitigating this difficulty. Specifically, we seek systems that reach orbital period values within the range of $P_{\rm orb}$ = 2-9 h as detached systems, for which donor stars become fully convective (MB is no longer acting), and subsequent orbital shrinkage occurs due to gravitational radiation. As a consequence of the orbital contraction, the donors may overflow their Roche lobes, initiating mass transfer once again. This leads to a significant reduction in the orbital period, i.e., $P_{\rm orb} \leq$ 1 hour, thus reaching the UCXB state. We will refer to this UCXB formation mechanism as {\it{channel~1}}. Additionally, the UCXB state can also be reached by systems that suffer a continuous, long-standing mass transfer episode, while its orbital period falls monotonically. Since the donor stars do not detach from their Roche lobe, the systems cannot be observed as binary pulsars. This UCXB formation mechanism constitutes the {\it{channel~2}}}.

%The second mass transfer episode in the system with $P_{\rm orb,i} = 1.68$~d under MB3 begins when the mass of the donor star starts to decrease again after the orbital period drops. At this moment, the donor star is in a degenerate state, and there is only angular momentum loss due to gravitational radiation. Since the radius of degenerate stars increases in response to mass loss, the onset of the RLOF results in high values for the MTR, thus leading to small mass ratio values ($\simeq 0.1$). This, in turn, results in an increase in the orbital semi-major axis, which dominates over the effects of gravitational radiation.

\referee{In Fig.~\ref{ucxb} we show the evolution of selected systems that reach the UCXB state throughout the channels mentioned above. The black stars in the figure indicate the start of mass transfer, and the red circles its end. We found that the range of orbital periods leading to the formation of BPs with $P_{\rm orb}$= 2-9~h and UCXBs depends on the MB prescription considered: }
%{\sout{As can be noticed, none of the systems evolved under MB0 produce BPs with $P_{\rm orb}$= 2-9~h, but there are UCXBs produced via {\it{channel~2}} resulting from systems with $P_{\rm orb,i}$ between 0.6 and 0.8~d. With MB2 we found BPs with $P_{\rm orb}$= 2-9~h from systems with $P_{\rm orb,i}=$ 0.45, 0.50, 0.55, and 0.90~d. There are also UCXBs produced via both {\it{channel~1}} and {\it{channel~2}}, with $P_{\rm orb,i}$ between 0.55 and 0.85~d. The MB3 prescription gives the largest number of UCXBs, some of them from {\it{channel~1}} and others from {\it{channel~2}}. The $P_{\rm orb,i}$ that gives rise to these UCXBs ranges from 0.60 to 2.29~d. There are also BPs with $P_{\rm orb}$= 2-9~h with $P_{\rm orb,i}$ from 0.45 to 0.60~d and from 0.60 to 2.52~d. Finally, CARB MB allows the formation of BPs with $P_{\rm orb}$= 2-9~h in systems with $P_{\rm orb,i}$ from 0.45 to 0.60~d and from 0.9 to 1.0~d; and give rise to UCXBs with $P_{\rm orb,i}$ from 0.70 to 0.90~d.} In all cases (i.e., with the four prescriptions) BPs with $P_{\rm orb}$= 2-9~h  have donor star masses $< 0.2~{M_{\odot}}$. 

\referee{
\begin{itemize}
    \item {Formation of ELM WDs BMPs with $P_{\rm orb}$= 2-9~h:}
    \begin{itemize}
        \item MB0: none
        \item MB2: $0.45~\rm \leq P_{\rm orb,i} \leq 0.55~d$, and $P_{\rm orb,i}=0.90~\rm d$
        \item MB3: $0.45~\rm \leq P_{\rm orb,i} \leq 0.6~d$ and $1.07~\rm \leq P_{\rm orb,i} \leq 2.52~\rm d$
        \item CARB MB: $0.45~\rm \leq P_{\rm orb,i} \leq 0.6~d$ and $0.9~\rm \leq P_{\rm orb,i} \leq 1.0~d$
    \end{itemize}
     \item {Formation of UCXBs:}
    \begin{itemize}
        \item MB0: $0.6~\rm d \leq P_{\rm orb,i} \leq 0.8~d$ ({\it{channel~2}})
        \item MB2: $0.55~\rm d \leq P_{\rm orb,i} \leq 0.85~d$ ({\it{channels~1}} and {\it{2}})
        \item MB3: $0.6~\rm d \leq P_{\rm orb,i} \leq 2.29~d$ ({\it{channels~1}} and {\it{2}})
        \item CARB MB: $0.7~\rm d \leq P_{\rm orb,i} \leq 0.9~d$ ({\it{channels~1}} and {\it{2}})
    \end{itemize}
\end{itemize}
}

\referee{Overall, when studying the formation of UCXBs, the choice of a MB prescription is of key relevance. If MB0 and CARB MB are considered, the range of $P_{\rm orb, i}$ that give rise to UCXBs is only 0.2~d, while for MB2 it is 0.3~d. However, MB3 prescription provides a window of initial orbital periods one order of magnitude wider than the other MBs (1.69~d). These results can be interpreted as evidence that at short $P_{\rm orb, i}$, a strong MB is necessary to account for the existence of UCXBs.}

%\referee{Overall, one remarkable finding is that with the MB3 prescription, the window in $P_{\rm orb, i}$ that provides models evolving to BP and then to UCXB widens when compared with the Skumanich law (MB0). As calculated in \citet{2019BAAA...61...87E}, when MB0 is used, the window in $P_{\rm orb,i}$ to reach the UCXB configuration via {\it{channel~1}} is $\simeq 0.05~$d width, while here we show it is $\simeq 1.25~$d if MB3 is considered. }

%------------------------------------------------------------------------------------------------------
\section{Comparison with the results available in the literature} \label{sec:comparison}

Various studies have been published to investigate the evolution of LMXBs and the formation of BPs, employing some of the novel MB prescriptions introduced in Van19 and \citet{2019ApJ...886L..31V}. \citet{2019cwdb.confE..13R} (hereafter Romero19) and \citet{2021MNRAS.503.3540C} (hereafter Chen21) studied the effects of MB0, MB2 and MB3; \citet{2021ApJ...909..174D} (hereafter Deng21) analysed the MB0, MB2, and CARB~MB; and \citet{2021MNRAS.506.3266S} (hereafter Soethe21) studied the CARB~MB. They use the stellar evolution code {\tt MESA}. Their results shed light on the topic, showing that the evolutionary paths of LMXBs are very dependent on the MB law. \referee{However, these models present differences between them}. 

As explicitly stated by Chen 21, the Intermediate MB prescription (MB3) has led to a variety of results. On one hand, Van19 were able to calculate sequences with $P_{\rm {orb,i}}$ up to $\simeq 400$~d, and found extremely high MTR values ($\simeq 10^{-2}~M_{\odot}~ {\rm yr}^{-1}$). Besides, they found all systems with low-mass donors end with $P_{\rm orb}$ smaller than the initial value. \referee{On the other hand, Chen21 calculated the evolution of systems with $P_{\rm orb,i}$ up to $\sim$ 25~d. They stopped the calculations since MTR reaches values up to $\sim 10^{-4}~M_{\odot}~{\rm yr}^{-1}$. They argue this large MTR value would lead the system to a common envelope phase. Furthermore, they found similar results for different values of the initial donor star mass.} As a consequence, they judged it is not possible to account for wide-orbit binaries, so they concluded MB3 is not viable as a universal MB law. On the contrary, Romero19 calculated the evolution of low-mass binaries with $P_{\rm orb,i}$ up to $\simeq$ 120~d. Unlike Van19, they found all systems with $P_{\rm orb,i} \gtrsim$ 25~d end their evolution with $P_{\rm orb}>P_{\rm orb,i}$, so they can also reproduce wide-orbit binaries. Chen21 and Romero19 found that the fine-tuning problem in producing UCXBs mitigates with this MB prescription.

The discrepancies among the results presented in Van19, Romero19, and Chen21 have been one of the motivations for performing the present study. Considering that our code is entirely independent of the one employed in the referred papers, we judge it is a good tool to disentangle this paradoxical situation by making careful comparisons between our and their results. 

As already discussed in Section~3, systems under MB3 suffer MTR as high as $10^{-4}~M_{\odot}~ {\rm yr}^{-1}$ when $P_{\rm orb,i}\gtrsim$ 19~d, and MTR is even higher for larger values of $P_{\rm orb,i}$, reaching values $\lesssim 10^{-2}~M_{\odot}~ {\rm yr}^{-1}$. When comparing with Van19 and Romero21, who have both calculated systems with larger $P_{\rm orb,i}$ than Chen21, we find that our results are in agreement (disagreement) with Van19 (Romero21), since all our simulated systems under MB3 shrink. Besides, we found MTR values compatible with Van19 (see Fig.~4 in their publication). \referee{Concerning} UCXB formation, the mitigation of the fine-tuning problem we found agrees with Romero21 and Chen21. 

Regarding the MB0 and MB2 prescriptions, \referee{our results generally agree with those given in Van19. Specifically, from the inspection of Fig.~2 (for MB0) and Fig.~3 (for MB2) in their publication, we notice that the MTR in our simulations reaches values consistent with theirs. Upon closer inspection of the upper-right corner in these figures, it becomes apparent that} the region around the bifurcation period $P_{\rm bif}$\footnote{\referee{The bifurcation period $P_{\rm bif}$ is defined as the value of $P_{\rm orb}$ for which the initial and final orbital periods take the same value.}}, which is $\simeq 1$~d, is more densely populated by tracks in the MB2 case than in the MB0. \referee{This result is in agreement with our simulations, as can be seen in Fig.~\ref{mtr}}. \referee{Romero19 models evolved under MB2 also populate the region around $P_{\rm bif}$ ($\sim 10$~d for these authors) more efficiently than with MB0}. \referee{Concerning} UCXB formation with MB2, our results agree with Chen21 and Romero19.

\begin{figure} 
 \centering
  \includegraphics[width=0.5\textwidth]{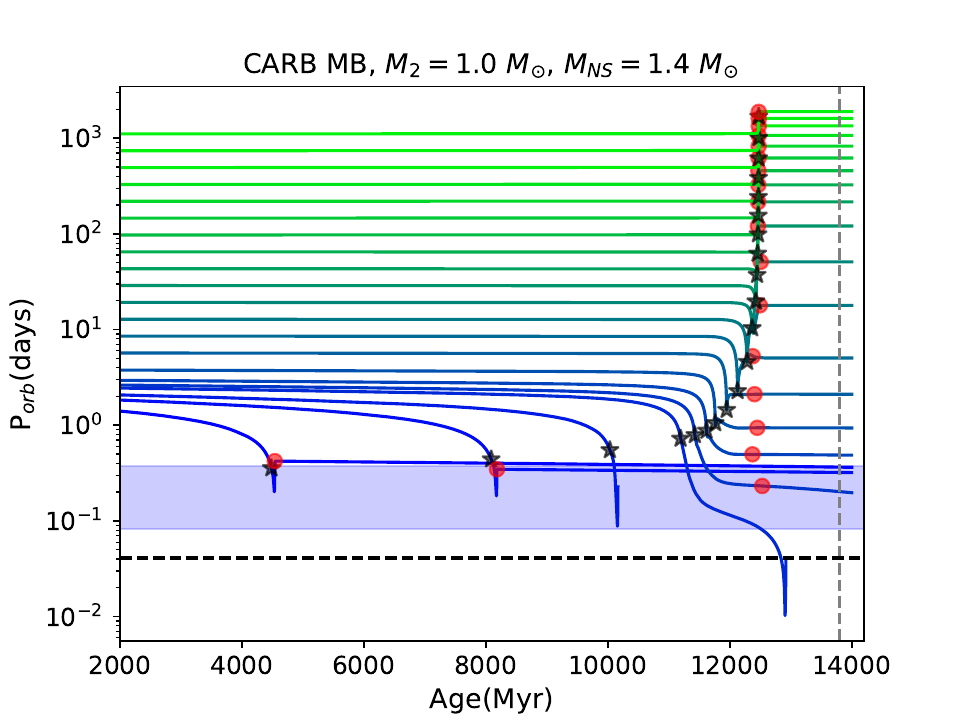}
  \caption{Orbital period as a function of time for systems under CARB MB, $M_{\rm 2,i}= 1.0~M_{\odot}$ and $M_{\rm NS,i}= 1.4~M_{\odot}$. Black stars indicate the start of mass transfer, and red circles its end. The horizontal light-blue band limits the region of orbital periods between 2 and 9~h. The horizontal black dashed line corresponds to 1~h. The vertical grey dashed line represents the age of the Universe.}
  \label{EPsoethe}
\end{figure}

\referee{To compare our results using the CARB~MB prescription against the results of Deng21 and Soethe21, we performed another set of calculations using initial values of $M_2= 1.0~M_{\odot}$ and  $M_{\rm NS}= 1.4~M_{\odot}$ (Fig.~\ref{EPsoethe}). In our simulations, $P_{\rm bif} \sim 10$~d, in agreement with Deng21 and in disagreement with Soethe21, who finds $P_{\rm bif} \sim 22$~d. Besides, Deng21 found that CARB~MB is more effective in producing BPs with orbital periods 0.1~d $< P_{\rm orb} < $1~d. This is in agreement with our findings and with Soethe21 (see their Fig.~3).}

%Regarding UCXB formation, we did not find any system formed via {\it{channel 1}}. This can be inferred from Fig.~\ref{EPsoethe}, which shows evolutionary tracks with $P_{\rm orb,i}= 1.68, 2.0, 2.2, 2.53, 2.7,$ and $ 3.0$~d together with the start and end of mass transfer. Nevertheless, we can not discard the possibility of still finding some UCXB via {\it{channel 1}} if a finer exploration in $P_{\rm orb,i}$ values were performed. On the other hand, Deng21 found one UCXB via this channel while Soethe21 found that systems with $P_{\rm orb,i}$ between 2.7 and 3.0~d evolve to UCXBs via {\it{channel 1}} (see their Fig. 3).

%Apart from this, \citet{2021MNRAS.506.3266S} intentionally looks for BPs with companions with mass values $< 0.26~M_{\odot}$ and finds systems  we see our results agree 

%\referee{With respect to the formation of BPs with $P_{\rm orb}$= 2-9~h, despite the fine-tuning problem for their formation mitigates, \citet{2021MNRAS.506.3266S},?? NO HABLA EN PARTICULAR DE ESTOS SISTEMAS. CAPAZ MEJOR NO DECIR NADA. ELLOS HABLAN DE BP CON COMPAÑERAS CON MASA < 0.26 }

\referee{We should state that the differences in the results available in the literature show quantitative and even qualitative disparities despite all these works were done with the same numerical tool. It may be possible that these differences were due to different choices of the input parameters necessary to carry out the simulations. On the other hand, we present a set of results obtained with an independent tool.}

\referee{It is remarkable that Van19 found that MB3 prescription seems to mitigate two discrepancies between observation and theory: the population of tracks around $P_{\rm orb} \sim 1$~d and the fine-tuning in the formation of UCXBs. This is in nice agreement with the results we presented in Section~\ref{sec:results}.}
 
%------------------------------------------------------------------------------------------------------
\section{Summary and Conclusions} \label{sec:Disc-Concl}

We have revisited the evolution of binary systems composed of a normal donor star and a neutron star. When these systems undergo mass transfer, they can be observed as LMXBs or UCXBs; when detached, they can be detected as binary pulsar systems. 

It is well known that these systems suffer some processes that lead to angular momentum losses (AML). Among AML are those due to gravitational radiation, non-conservative mass transfer, and magnetic braking. At present, the latter is considered rather uncertain, and because of this reason, some different prescriptions have been recently proposed in the literature. This is the case of the MB prescriptions MB2, MB3, and CARB~MB from \citet{2019MNRAS.483.5595V} (Van19) and \citet{2019ApJ...886L..31V}.

To study the effects of these three prescriptions together with the standard Skumanich law (MB0) on the evolution of LMXBs and UCXBs, we have considered systems with initial masses of 1.25~$M_{\odot}$ for the donor star and 1.30~$M_{\odot}$ for the accreting NS. We computed a wide range of initial orbital periods. In this way, we present a general view of the effects of the four MB prescriptions in a consistent way. The calculations were carried out using our stellar code \citep{2003MNRAS.342...50B} which is completely independent of {\tt MESA} \citep{2015ApJS..220...15P}, a code commonly employed in analogous computations.

We found that all these MB laws allow for the occurrence of systems with tight ($P_{\rm orb}< 1$~h) and wide ($P_{\rm orb}> 50$~d) orbits. For the cases of MB0, MB2, and CARB~MB we found orbital periods between $8\; \rm{min} \lesssim P_{\rm orb} \lesssim 8\; \rm{yr}$, while for MB3 the range is $6\; \rm{min}  \lesssim P_{\rm orb} \lesssim 55\; \rm{d}$. It is interesting to notice that when MB3 is used, it is possible to reach these $P_{\rm orb}$ values while having high ($< 10^{-2.27}~M_{\odot}~ {\rm yr}^{-1}$) MTR. The MB3 prescription deeply differentiates from MB0, MB2, and CARB~MB. That is, MB3 gives distinct results in all the features that we have examined in this work, i.e., the evolution of the orbital period, the MTR, and the final donor star and NS masses. Furthermore, MB0, MB2, and CARB~MB become almost indistinguishable from each other when the orbital period is large enough ($ P_{\rm orb,i} \gtrsim 50\; \rm{d}$).

The well-known fine-tuning problem in the formation of UCXBs is mitigated with the MB2 and CARB MB prescriptions, while MB3 proves to be the most effective law for resolving it. Similarly, the MB2 and CARB MB prescriptions help to populate the region in the $M_2 - P_{\rm orb}$ plane occupied by observed systems with $P_{\rm orb} \sim 1$ d, while MB3 is the most successful law for achieving this.
%\ale{This MB prescription also helps to populate the region in the $M_2 - P_{\rm orb}$ plane occupied by  systems whit $P_{\rm orb} \sim 1$ d.}

While the MB3 prescription is promising in connection with the fine-tuning problem of UCXB formation, we found that our set of models reaches orbital periods up to 55~days, which is not large enough to account for the systems observed with wider orbits (see Fig.~\ref{mtr}, panel (c)). We should remark that these results were found assuming a fixed pair of initial masses. A more extensive parameter exploration should be conducted to evaluate whether models with MB3 can reach such periods. This will be the subject of future work.

%\referee{(CARB~MB) They note that with this MB prescription, all persistent LMXBs can be reproduced and recommend its use instead of the Skumanich MB to model galactic and extragalactic LMXBs with NSs} ESTAS COSAS NO SE SI PONERLAS EN LA INTRO O EN LA DISCUSION.

%\section{OTRA SECCION TIPO TRABAJO FUTURO O APMLIAR EL TITULO DE LA SECCION 5}

%%%%%%%%%%%%%%%%%%%%%%%%%%%%%%%%%%%%%%%%%%%%%%%%%%
\section*{Data Availability}

The data generated by our numerical code are available from the
corresponding author upon request. All remaining data underlying
this article are available in the article and references therein.
%%%%%%%%%%%%%%%%%%%%%%%%%%%%%%%%%%%%%%%%%%%%%%%%%%
\section*{Acknowledgments}

We want to acknowledge our anonymous referee for her/his comments which were a valuable guide for improving the original version of this paper.



\bibliographystyle{mnras}
\bibliography{biblio} % if your bibtex file is called example.bib

%\begin{thebibliography}{99}
%\end{thebibliography}

%%%%%%%%%%%%%%%%%%%%%%%%%%%%%%%%%%%%%%%%%%%%%%%%%%

%%%%%%%%%%%%%%%%% APPENDICES %%%%%%%%%%%%%%%%%%%%%

%\appendix


%If you want to present additional material which would interrupt the flow of the main paper, it can be placed in an Appendix which appears after the list of references.

%\end{itemize}

%%%%%%%%%%%%%%%%%%%%%%%%%%%%%%%%%%%%%%%%%%%%%%%%%%

% Don't change these lines
\bsp	% typesetting comment
\label{lastpage}
\end{document}